\PassOptionsToPackage{dvipsnames}{xcolor}
\PassOptionsToPackage{hyphens}{url}
\documentclass[a4paper,onecolumn,11pt,unpublished]{quantumarticle}
\pdfoutput=1
\usepackage[utf8]{inputenc}
\usepackage[english]{babel}
\usepackage[T1]{fontenc}
\usepackage{amsmath}
\usepackage{hyperref}
\usepackage{physics}
\usepackage{subcaption}
\usepackage{float}
\usepackage{tikz}
\usepackage{lipsum}
\usepackage[ruled]{algorithm2e}
\usepackage{csquotes}
\usepackage{booktabs}
\usepackage{listings}
\usepackage{simpleicons}
\usepackage[numbers,sort&compress]{natbib}

\definecolor{codegreen}{rgb}{0,0.6,0}
\definecolor{codegray}{rgb}{0.5,0.5,0.5}
\definecolor{codepurple}{rgb}{0.58,0,0.82}
\definecolor{backcolour}{rgb}{0.95,0.95,0.92}

\lstdefinestyle{mystyle}{
    backgroundcolor=\color{backcolour},   
    commentstyle=\color{codegreen},
    keywordstyle=\color{magenta},
    numberstyle=\tiny\color{codegray},
    stringstyle=\color{codepurple},
    basicstyle=\ttfamily\footnotesize,
    breakatwhitespace=false,         
    breaklines=true,                 
    captionpos=b,                    
    keepspaces=true,                 
    numbers=left,                    
    numbersep=5pt,                  
    showspaces=false,                
    showstringspaces=false,
    showtabs=false,                  
    tabsize=2
}
\makeatletter
\newcommand{\urlcolornameref}[1]{%
  \begingroup
    \hypersetup{linkcolor=\@urlcolor}
    \nameref{#1}%
  \endgroup
}
\makeatother

\author[1,2]{Iyán Méndez Veiga}
\orcid{0009-0004-8249-4661}
\email{iyan.mendezveiga@hslu.ch}
\author[1]{Esther Hänggi}
\orcid{0000-0002-9760-8535}
\affil[1]{Lucerne University of Applied Sciences and Arts, 6343 Rotkreuz, Switzerland}
\affil[2]{Institute for Theoretical Physics, ETH Zurich, 8093 Zurich, Switzerland}

\begin{document}

\title{Reproducible Builds for Quantum Computing}
\maketitle

\begin{abstract}
Reproducible builds are a set of software development practices that establish an independently verifiable path from source code to binary artifacts, helping to detect and mitigate certain classes of supply chain attacks. Although quantum computing is a rapidly evolving field of research, it can already benefit from adopting reproducible builds. This paper aims to bridge the gap between the quantum computing and reproducible builds communities. We propose a generalization of the definition of reproducible builds in the quantum setting, motivated by two threat models: one targeting the confidentiality of end users' data during circuit preparation and submission to a quantum computer, and another compromising the integrity of quantum computation results. This work presents three examples that show how classical information can be hidden in transpiled quantum circuits, and two cases illustrating how even minimal modifications to these circuits can lead to incorrect quantum computation results. Our work provides initial steps towards a framework for reproducibility in quantum software toolchains.
\end{abstract}

\section{Introduction}
Quantum computing is a very active field of research. In the 90s, theoretical work demonstrated potential speedups for specific computational problems compared to classical algorithms~\cite{shor1,grover,shor2}, but practical implementation is still at an early stage. The field has nevertheless evolved rapidly with large public and private investments \cite{mckinsey2025,mckinsey2025quantumtechnology,EU-quantum-flagship,nqi2024budget} and ambitious roadmaps~\cite{IBM-roadmap-2025,Google-QuantumAI,Quantinuum-roadmap}. The first software development kits (SDK) that interact with these novel machines are already available (see e.g.,~\cite{qiskit2024,amazon-braket-python-sdk}).

At the same time, several established best practices and standards from classical computing have not yet been fully translated into the quantum computing context. In particular, this paper extends the concept of reproducible builds to current quantum computing workflows, with the aim of protecting users from specific threat vectors targeting confidentiality and integrity.

The structure of the paper is as follows. Sections~\ref{sec:r-b} and~\ref{sec:qc} introduce the concepts of reproducible builds and quantum computing, respectively. Given the goal of bridging the gap between the reproducible builds and quantum computing communities, these introductory sections provide the necessary background. Readers already familiar with either topic may wish to skip the corresponding section. Section~\ref{sec:r-b-q} presents a practical definition of reproducible builds tailored for current quantum computing workflows. Section~\ref{sec:threat_models} introduces the two threat models considered throughout the remainder of the paper. Sections~\ref{sec:attacks-confidentiality} and~\ref{sec:attacks-integrity} detail examples of threat vectors that exploit the non-reproducibility of transpiled quantum circuits, affecting both the confidentiality of user data and the integrity of computation results. These examples are not directly exploitable since they rely on authenticated access to data exchanged with the quantum cloud platform, but they show the usefulness of reproducible builds in the quantum setting for risk mitigation. Section~\ref{sec:ibm-r-b} identifies what causes Qiskit~\cite{qiskit2024} to produce non-reproducible circuits and proposes specific changes to enable reproducible transpilation. Finally, Section~\ref{sec:conclusion} summarizes the main contributions and outlines potential directions for future research.

\section{Reproducible Builds for Quantum Engineers}\label{sec:r-b}
The vast majority of modern software is not written in a language that computers can understand directly i.e., machine code. Instead, programming languages are used. There are thousands of different programming languages~\cite{wiki:List_of_programming_languages}, but they have something in common: they can be easily read and written by humans. The description of software in any of these human-friendly languages is referred to as source code. The conversion between source code to machine code is done by an interpreter or a compiler~\cite{dragoonBook}. The challenging task of converting machine code to source code is known as reverse engineering~\cite{eilam2011reversing}.

End users only need the final machine code to execute the software. If, in addition, the source code is provided, this software is called~\emph{open source}\footnote{If some further freedoms are guaranteed (e.g., the freedom to modify the software and redistribute it), it is called free software, or \emph{libre} software~\cite{stallman2002free} to emphasize that is a matter of liberty, not price. Both classes of software are referred together as Free and Open-Source Software (FOSS).}~\cite{opensource}.

\begin{figure}[ht]
\centering
\includegraphics[width=0.4\linewidth]{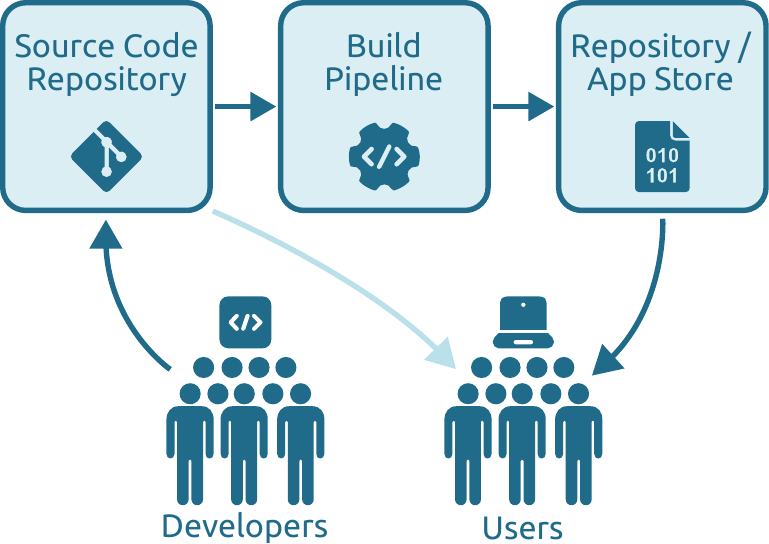}
\caption{\textbf{Illustration of a typical open-source software development workflow.} Developers contribute by committing changes to a shared source code repository. When ready, they create a new release by tagging a specific version of the code, which triggers an automated build pipeline. This process generates the final executable or binary files intended for end users. These files are then distributed through package repositories or app stores.}
\label{fig:diagram_open_source}
\end{figure}

It is important to note that, even if the source code is available, most end users will not compile it themselves and will, instead, rely on machine code provided by developers or by some other trusted party (e.g., an app store), as shown in Figure~\ref{fig:diagram_open_source}. The reason for this is that the compilation process for large software projects can be complex and computationally intensive (e.g., compiling modern web browsers can take several hours on high-end PCs). However, in general, there is no easy way to prove that these binaries were obtained from that particular source code.

While it is extremely difficult to audit machine code, source code can be audited to find bugs and backdoors\footnote{A backdoor is a piece of code intended to change the behavior of a software in a malicious way, e.g., to allow an attacker to execute remote commands.}. However, even if the source code is audited, the translation to machine code can introduce further modifications that are not directly visible in the source code alone. In the worst possible scenario, this translation can be controlled by an attacker (see, e.g.,~\cite{CVE-2024-3094}) to introduce malicious changes. This is a particular type of \emph{supply chain attack}~\cite{supply-chain-attacks}, a cyberattack that targets the trusted tools, software, or services used to build software, in order to compromise the final artifact or its users. For example, the safe source code of a web browser can be modified during compilation to include a keylogger that sends passwords typed by their users to a remote attacker. This modification is completely unnoticeable by auditors that focus solely on the source code, but it will affect most users since, as mentioned above, they rarely compile the software themselves.

The solution to this problem is to establish a verifiable connection between source code and machine code. Since reverse engineering machine code back to source is not feasible for large software, and hardly practical for small software, this connection can only be ensured in the opposite direction i.e., by verifying that the same source consistently produces the same machine code. When the same source code, transformed in a similar\footnote{By \emph{similar} it should be understood, among other things that will be covered in more detail later, that the exact same versions of the building tools are used.} environment, produces bit-by-bit identical machine code, the software is said to have a deterministic build, or simply, that the software is reproducible~\cite{r-b-paper}. Reproducible builds are, therefore, a set of software development practices that create an independently verifiable path from source to machine code~\cite{r-b-web}.

Reproducible builds enables users willing to build software from source code to take the role of verifiers and check that the official distributed binaries indeed stem from the provided source code. If they all obtain exactly the same final binaries, then the link from machine code to source code is confirmed. Otherwise, it is called non-reproducible and no statement about the link between machine code and source code can be made. Non-reproducible software can be due to a) some harmless non-determinism in the source code or building pipeline (e.g., timestamps or random seeds determined at build time), or b) an alteration, which might be malicious or not. Since it is not possible to distinguish between these cases, it is unjustified to trust non-reproducible software.

The bit-by-bit identical machine code is a stringent but necessary requirement since a one-bit difference can be enough to enable an attack (see, e.g.,~\cite{CVE-2002-0083}).  If a software is reproducible and a difference is observed by a verifier in the machine code, even just a single bit, then it is uncertain whether the secure and audited path from the source code to the final machine code was followed, and those binaries should not be trusted. It is important to emphasize that reproducible builds should not be seen as a replacement for source code audits: insecure software will remain insecure even if it is modified to be reproducible.

Reproducible builds alone do not bring any benefits to end users. Security benefits only arise with additional infrastructure including first, a set of independent verifiers that build source code and compare their builds with the distributed machine code~\cite{r-b-hslu}; and second, package managers and app stores that use the output of these verifiers to warn users, or even block the distribution of machine code that is not reproducible. Only then, reproducible builds can detect and prevent this kind of supply chain attacks. While many relevant software is already reproducible (e.g., Chromium~\cite{chromium-deterministic-builds}, Bitcoin wallet~\cite{bitcoin-deterministic-builds}, GrapheneOS~\cite{grapheneos-reproducible-builds}), only very few independent verifiers exist (see e.g., \cite{archlinux-rebuilders}). To the best of our knowledge, reproducibility status is not yet leveraged by package managers or app stores, presenting an opportunity for future improvements in transparency and user trust. Nevertheless, reproducibility has lately been elevated to a recognized best practice in secure software supply chains, with frameworks such as Supply-chain Levels for Software Artifacts (SLSA) explicitly including verified reproducible builds as a requirement at the highest assurance level~\cite{slsa}.

In summary, supply chain attacks that compromise the building process of software can be very tricky to detect by simply inspecting the final binary artifacts. Reproducible builds can ensure that bit-by-bit identical artifacts are obtained from the exact same source code.  Once bit-by-bit identical and deterministic builds are available, deviations detected by independent verifiers, if properly announced, can protect end users from this kind of supply chain attacks.

\section{Quantum Computing for the Reproducible Builds Community}\label{sec:qc}
Classical digital computers process and store information encoded in zeros and ones, i.e., bits. Physically, a bit can be implemented, for example, with a high voltage to represent a one, and with a low voltage to represent a zero. Modern digital computers can process trillions of these bits every second using integrated circuits with billions of transistors~\cite{1273999}. Digital quantum computers, on the other hand, do computations with quantum bits or qubits~\cite{Benioff1980}. The main difference between a bit and a qubit is that the minimum amount of quantum information can not only be in a state\footnote{A quantum state is the complete description of a quantum system, containing all that can be known about it e.g., the probabilities of different measurement outcomes. Mathematically, quantum states are represented as elements of a vector space. For qubits, this vector space has dimension two.} zero or one but also in any linear combination of both, a so-called \emph{superposition}. Mathematically, following the standard notation in quantum information theory~\cite{NieChu00} where the state zero is represented with $\ket{0}$ and state one with $\ket{1}$, a qubit can be in the following general state
\begin{equation*}
\ket{\text{qubit}} = \alpha \ket{0} + \beta \ket{1}\,,
\end{equation*}
where the coefficients $\alpha$ and $\beta$ are any complex numbers such that $|\alpha|^2+|\beta|^2=1$.

When qubits are \emph{measured} (read) they can only take two possible values, $\ket{0}$ or $\ket{1}$, i.e., qubits \emph{collapse} into bits when \emph{observed}. This is another key difference from classical computers: quantum computers, like the underlying theory they are based on, are intrinsically probabilistic. The likelihood that a qubit collapses to $\ket{0}$ or $\ket{1}$ upon measurement is determined by the squared magnitudes of its complex amplitudes, specifically $\Pr[\ket{0}] = |\alpha|^2$ and $\Pr[\ket{1}] = |\beta|^2$.

Qubits are a generalization of bits. In particular, bits zero and one can be encoded in a qubit by choosing $\alpha=1$ and $\beta=0$, and $\alpha=0$ and $\beta=1$, respectively. Physically, a qubit can be implemented with a quantum two-level system. Trapped ions, superconducting circuits, or Nitrogen-Vacancy centers in nanodiamonds are examples of current physical platforms used to encode qubits~\cite{Chae2024}.

An ideal qubit platform for building quantum computers should be able to store, manipulate, and read out quantum information reliably, while scaling to systems large enough to solve meaningful problems. Qubits must be well isolated from noise and at the same time remain controllable. Unlike in classical computing, where the dominant hardware technologies are well established, it is still not clear which physical platform will prove most suitable for building quantum computers. Each of the platforms mentioned before has its advantages and disadvantages, and perhaps in the future completely different physical systems will be used. Finding a platform with all the desired properties is an active field of research~\cite{Chae2024}.

Quantum computers can perform operations with qubits that are in a superposition, transforming them into different superpositions. This is the foundation of what is known as \emph{quantum parallelism}, a key feature that many quantum algorithms exploit. Roughly speaking, quantum parallelism allows computing a function $f(x)$ on a superposition with different values of $x$ to obtain a superposition of the individual computations. It is important to note that this is not the same as being able to compute and access all values of $f(x)$ at once.  By computing a function on all possible inputs in superposition and measuring the results, one would simply obtain the result corresponding to one randomly chosen input. This does not lead to a speedup. The challenge is, therefore, to create an \emph{interference}\footnote{States in a superposition can interfere with each other and this can lead to reinforcing some outcomes (constructive interference) and suppressing others (destructive interference) when measured.} between the results corresponding to different inputs such that the probability to obtain an interesting result is enhanced.

The field of quantum computing was born in the 80s. Some scientists, among them Richard P.~Feynman, soon realized that simulating quantum systems using classical digital computers was a hard task, a problem that would require exponentially more resources and computation power to simulate just slightly larger quantum systems~\cite{preskill-history-qc}. They also noticed that perhaps a computer based on the principles of quantum mechanics, a quantum computer, would make this task feasible. This motivated a lot of theoretical work and by the middle of the following decade the first quantum algorithms were discovered: Deutsch–Jozsa's~\cite{deutsch-jozsa}, Bernstein–Vazirani's~\cite{bernstein-vazirani} and Simon's~\cite{simon} algorithms were the first demonstrations of how using qubits could have an advantage over classical bits. However, the two algorithms that finally spiked interest into quantum computing and started a race to build a quantum computer were Shor's algorithm~\cite{shor1} in 1994 and Grover's algorithm~\cite{grover} in 1996.

Shor's algorithm allows a quantum computer to \emph{efficiently}\footnote{In computational complexity theory, \emph{efficiently} means solvable in polynomial time with respect to the input size~\cite{Arora_Barak_2009}.} solve three problems: the integer factorization problem (given a large integer, determining its prime factors), the discrete logarithm problem (given a group element and a base, finding the exponent that produces it), and the period-finding problem (given a function, determining its period). The best known classical algorithm for finding the prime factors of an integer, the general number field sieve~\cite{number-field-sieve}, only works in sub-exponential time\footnote{An algorithm is said to work in sub-exponential time if its runtime is smaller than exponential but it is not bounded by any polynomial. This distinction is crucial because it separates practically feasible algorithms from others that quickly become intractable.} while Shor's algorithm runs in polynomial time. One of the reasons this algorithm raised a lot of interest and concern is because solving these problems efficiently would mean breaking widely used public-key cryptography\footnote{This is the reason why there is a transition to so-called post-quantum cryptography, where public-key cryptography algorithms such the asymmetric cipher RSA-OAEP~\cite{RSA,RSA-OAEP} or the digital signature ECDSA~\cite{ECDSA} are being replaced with new quantum-safe algorithms, such as ML-KEM~\cite{ML-KEM} or ML-DSA~\cite{ML-DSA}, that rely on problems that are believed not to be efficiently solvable by classical and quantum computers.}.

Grover's algorithm is a quantum algorithm for searching an unstructured database or solving unstructured search problems. It is more general than Shor's algorithm in the sense that it can be applied to many different problems, but it only provides a quadratic speedup with respect to the analogous classical search. This means that problems with exponential-time complexity remain so. This algorithm also impacts cryptography, though its consequences are less severe than those of Shor’s algorithm. For example, the exhaustive search of a 128-bit secret key in a symmetric block cipher could be done in roughly $2^{64}$ iterations rather than $2^{128}$ using Grover's algorithm. Doubling the key size, a simpler task compared to coming up with completely new algorithms, compensates for Grover's quadratic speedup by requiring again the same number of steps as before.

Quantum computers already exist, but none are yet capable of executing Shor’s or Grover’s algorithms on large inputs. The main reason is that these devices still contain a limited number of qubits (they lack scalability) and quantum information cannot be processed and stored reliably for long periods of time. Building a quantum computer is an engineering problem, but a very difficult one. Many solutions developed in the early years of classical computing cannot be applied directly to quantum computers and must be adapted (e.g., error correction). Other problems are completely novel (e.g., cooling down electronic circuits with millikelvin cryogenics~\cite{Hollister2022}) and some problems may only be discovered when trying to further scale current devices.

The timeline for the development of large-scale, fault-tolerant quantum computers capable of running algorithms such as Shor's remains uncertain. Estimates vary widely, with some researchers projecting breakthroughs within a decade~\cite{quantum-threat-2024}, while others remain more cautious or skeptical~\cite{gent2023}. Regardless of long-term predictions, recent advances in quantum hardware and software~\cite{IBM-2024,Google-quantum-error-correction-2024, AWS-2024} have enabled public access to noisy quantum devices with hundreds of qubits. This paper adopts a practical perspective: given that such systems are already in use, it examines how well-established techniques from classical computing, particularly those related to security and reproducibility, can be adapted to enhance the reliability and trustworthiness of current quantum computing workflows.

\section{Reproducible Builds for Quantum Computing}\label{sec:r-b-q}

Even though Section \ref{sec:r-b} focused solely on software, the concepts of source and machine code can be generalized to include other kinds of artifacts. For example, the source code of this paper written with \LaTeX{} are the~\texttt{.tex} and~\texttt{.bib} files, while the ``machine code'' is the compiled~\texttt{.pdf}, even though this is never executed by a computer, but rather read and interpreted by a different program, e.g., a PDF reader. Similarly, configurations or scripts can be transformed, rather than compiled, to more lengthy and complex text files using software. In these cases, text files originally written by people are referred to as source code, and the transformed ones as ``machine code'' or artifacts.

Current quantum ``software'' consists of a classical description of the circuit that the quantum computer should execute, rather than being represented as quantum information (e.g., stored in qubits). This may change in the future, when reliable quantum memories are available and can be used to store both quantum states and algorithms to transform them. At present, quantum computations are expressed as quantum circuits, which are a sequence of different instructions that prepare, transform and measure qubits. As with classical circuits, transformations are implemented using logic gates, and only a small set of quantum logic gates, a so-called set of universal quantum gates~\cite{NieChu00}, is sufficient to describe any possible quantum computation.

\begin{figure}[ht]
\centering
\includegraphics[width=0.5\linewidth]{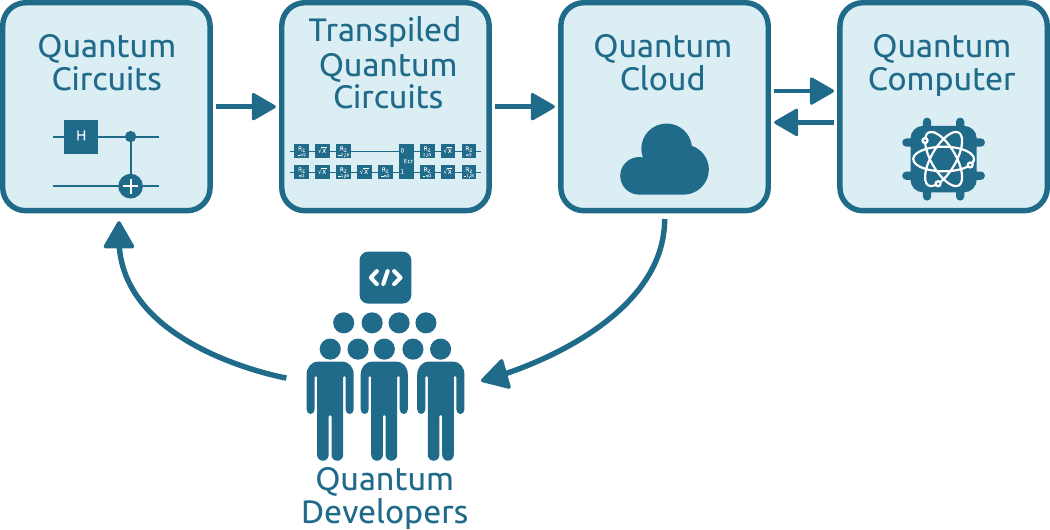}
\caption{\textbf{Illustration of the current quantum computing workflow.} Quantum developers implement algorithms as quantum circuits using an SDK such as Qiskit~\cite{qiskit2024}. These circuits are transpiled (see Sec.~\ref{sec:transpilation}) and sent to a quantum cloud, which provides access to a quantum backend. The quantum computer executes the circuit, and the results are returned via the cloud platform to the developers for further analysis.}
\label{fig:diagram_quantum_computing}
\end{figure}

In the following, the usual terms used in (classical) reproducible builds to quantum computing are generalized.

\paragraph{Source code} The source code of quantum software is the quantum circuit that fully describes the computation, as it is originally written (i.e., typed into a computer) by a human in plain text (i.e., human readable alphanumeric characters). In other words, it is the set of all files that contain human-readable instructions written in a programming language that can be understood by programmers that uniquely describe a quantum circuit. Examples of quantum source code include: a quantum circuit described using the Qiskit SDK~\cite{qiskit2024}, a circuit described using Amazon Braket SDK~\cite{amazon-braket-python-sdk}, or a quantum circuit and an algorithm written using the Open Quantum Assembly Language (OpenQASM)~\cite{OpenQASM3}.

\paragraph{Quantum build toolchain} The quantum build toolchain is the set of tools required to transform the source code of a quantum circuit into code (binary or not) that can be executed by a quantum processing unit (QPU). Similar to classical software, and depending on the type of programming language used to describe the circuit, this set may include a compiler, an assembler, a linker, an interpreter, etc. Current quantum build toolchains contain an additional tool: the \emph{transpiler}, which transforms a quantum circuit into an equivalent one that can be executed on the target quantum hardware. Some additional tools are left out of the scope of the quantum build toolchain e.g., the backend software that transforms the quantum (transpiled) circuits into actual operations on the physical qubits (e.g., microwave pulses). Although this exclusion may initially appear unjustified, an analogous distinction is made in classical software. For example, CPU microcode is left out of the scope of reproducible builds, even though it has the potential to change the runtime behavior of any software.

\paragraph{Quantum artifacts} In addition to the classical artifacts i.e., any output of a (classical) build toolchain with the exception of plaintext logs (e.g., executables, libraries, documentation, etc.), quantum artifacts are defined as the code (binary or not) that is submitted via a quantum cloud platform to a QPU for execution. For example, when using IBM's quantum backends, the quantum artifact is the binary payload containing the transpiled circuit. It is important to emphasize that in current quantum computing workflows (see Figure~\ref{fig:diagram_quantum_computing}) these quantum artifacts only contain classical information. However, it might be possible that in the future, when reliable quantum memories and quantum communication channels are available, quantum states are also distributed as part of the quantum artifacts. These states could be used e.g., as the initial state for a quantum computation.

\paragraph{Reproducible quantum builds} Quantum software is said to be reproducible if, after designating a specific version of its source code as well as the versions of all the tools from the build toolchain, every build produces bit-by-bit identical classical and quantum artifacts. In other words, in current quantum computing workflows, reproducible quantum builds mean that an independently verifiable path exists from the source code, i.e., the description of the quantum circuits as originally written by a human, to the binary artifacts, i.e., the payload containing transpiled quantum circuits.\footnote{This definition does not imply that the outputs of executing a quantum circuit on a quantum computer must be deterministic.} Later, the terms \emph{reproducible quantum builds} and \emph{reproducible transpiled (quantum) circuits} are used interchangeably.

This definition should be revised if, in the future, quantum states are distributed as part of the quantum artifacts, since in that case the bit-by-bit requirement is incompatible with the distribution of quantum information. 

\subsection{Transpilation}\label{sec:transpilation}

Given that quantum artifacts currently consist of the transpiled quantum circuits, this section explains the transpilation process and its implementation in Qiskit. Transpilation is crucial since it is the source of unreproducibility in our threat models (Section \ref{sec:threat_models}) and that is precisely what will be exploited in all later examples (Sections \ref{sec:attacks-confidentiality} and \ref{sec:attacks-integrity}).

Transpilation is the process of obtaining a quantum circuit that can be executed on particular quantum hardware. For example, a small quantum circuit can be designed using three-qubit gates and assuming that all qubits are ``connected'', but the actual quantum hardware might only support a small subset of one- and two-qubit gates, and perhaps only certain qubits can interact with each other. Therefore, the transpiled circuit is one that implements the three-qubit gates as a composition of single and two-qubit gates, and it includes additional SWAP gates taking into account the connectivity of the physical qubits in the real hardware. Typically, a transpiled circuit is equivalent to the original one. However, in some cases, an approximately equivalent circuit may be generated instead e.g., to reduce the number of SWAP gates or the overall circuit depth. This trade-off can be preferable on current noisy quantum computers.

\begin{figure}[H]
\centering
\includegraphics[width=0.9\linewidth]{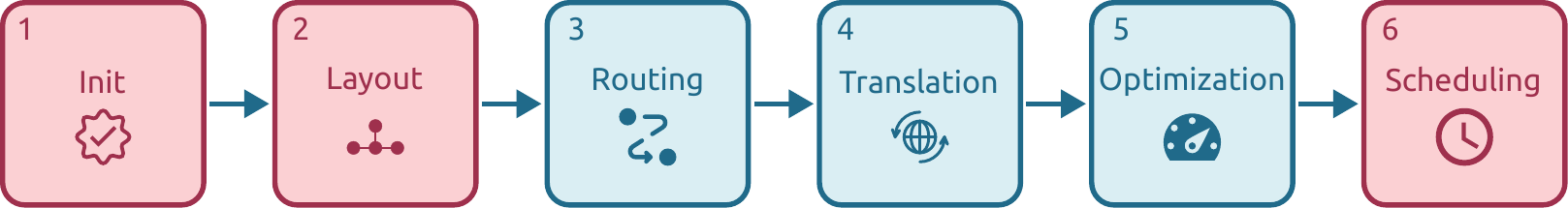}
\caption{\textbf{The six stages of the transpilation in Qiskit}~\cite{qiskit-transpiler-stages}. The stages highlighted in red are modified in Sections~\ref{sec:attacks-confidentiality} and~\ref{sec:attacks-integrity} to implement our examples.}
\label{fig:transpilation}
\end{figure}

Transpilation is not performed in a single step, but rather in several sequential steps or stages. For example, Qiskit's default transpiler pipeline (see Figure~\ref{fig:transpilation}) consists of six stages~\cite{qiskit-transpiler-stages}:
\begin{enumerate}
    \item In the \textbf{init stage} initial checks are run and all instructions are converted into single- and two-qubit gates.
    \item In the \textbf{layout stage} the virtual qubits are mapped to the QPU's physical qubits.
    \item The \textbf{routing stage} inserts SWAP gates where needed taking into account the mapping from the previous stage and the QPU's connectivity.
    \item In the \textbf{translation stage} all the gates are implemented in terms of the QPU's basis set of instructions.
    \item In the \textbf{optimization stage} the goal is to obtain more efficient representations of the quantum circuit or approximations given a certain condition (e.g., a maximum depth).
    \item In the last stage, the \textbf{scheduling stage}, all the instructions are assigned a start time taking into account hardware timing constraints. Different instructions might take different amounts of time, so they all have to be properly time-aligned in order to correctly implement the designed circuit.
\end{enumerate}

The transpilation process can become a very complex and computationally demanding task, as well as having a critical impact on the results of the quantum computation.

\section{Threat models}\label{sec:threat_models}

The focus on threats in this section is primarily conceptual and serves to illustrate the security benefits that arise when reproducible quantum builds are employed. By formulating explicit threat models, this offers a framework for demonstrating how deterministic and verifiable transpilation processes strengthen the confidentiality and integrity of quantum computing workflows.

A threat model is a structured description of the security assumptions, potential adversaries, their capabilities, and the possible attack vectors relevant to a given system. Threat modeling~\cite{Schneier_threat_modeling} helps clarify what kinds of attacks are considered, which components are within the scope of the model, which ones are trusted, and where vulnerabilities may arise. Including explicit threat models is essential in order to reason clearly about security goals and to design appropriate countermeasures.

In this section, two threat models are introduced in the context of current quantum computing workflows (see Figure~\ref{fig:diagram_quantum_computing}): one targeting the confidentiality of users' data and the other targeting the integrity of quantum computations.

Transpilation plays a crucial role in both models. Currently, transpiled circuits are expected to differ even when given identical inputs, which creates opportunities for adversarial manipulation. An attacker can exploit the transpilation steps to introduce malicious modifications. These threat models motivate the need for reproducible quantum builds, since both attacks are easily detectable and mitigated if the transpilation process produces deterministic and verifiable artifacts.

\paragraph{Common assumptions}

Before presenting the individual threat models, it is useful to state a set of common assumptions that define the scope of the analysis and limit the attacker's capabilities. These assumptions clarify which parts of the system are considered trusted and which attacks are deemed out of scope. For example, if an adversary has full physical or remote access to the user's computer, they could trivially extract any information, making confidentiality impossible to guarantee. Such scenarios are therefore excluded. Instead, the focus is on attacks that exploit weaknesses within the quantum software toolchain and workflow under more realistic constraints, where certain components remain trusted and isolation boundaries are respected. Based on these considerations, the following common assumptions are made:
\begin{itemize}
    \item The classical computer used for circuit preparation and submission is not physically accessible to the attacker.
    \item All the software used in the quantum workflow (e.g., circuit generation, compilation, and transpilation) is in scope.
    \item The build and transpilation processes are performed offline. External libraries and dependencies can be fetched before the build and transpilation starts, and network communication is allowed again at the end of the workflow when the final quantum circuit is transmitted to the cloud backend. This is a common assumption and done in practice in classical software pipelines (see e.g., Debian~\cite{debian-build-policy} and RHEL/Fedora~\cite{fedora-build-policy}).
    \item The user does not intentionally collaborate with the attacker.
\end{itemize}

\subsection{Confidentiality}\label{sec:threat_model_confidentiality}

This threat model focuses on attacks that compromise the \emph{confidentiality} of user data during the preparation and submission of quantum circuits. Confidentiality means ensuring that sensitive information is not disclosed to unauthorized parties. In Section~\ref{sec:attacks-confidentiality}, it will be shown how an attacker can modify the transpilation process to covertly extract sensitive classical information from the user's local environment and embed it into the final circuit sent to the quantum cloud. To clearly define the scope of such attacks and how they can be mitigated through reproducible transpilation, the assumptions, attacker capabilities, and threat vectors are specified precisely.

\begin{figure}[ht]
\centering
\includegraphics[width=0.8\linewidth]{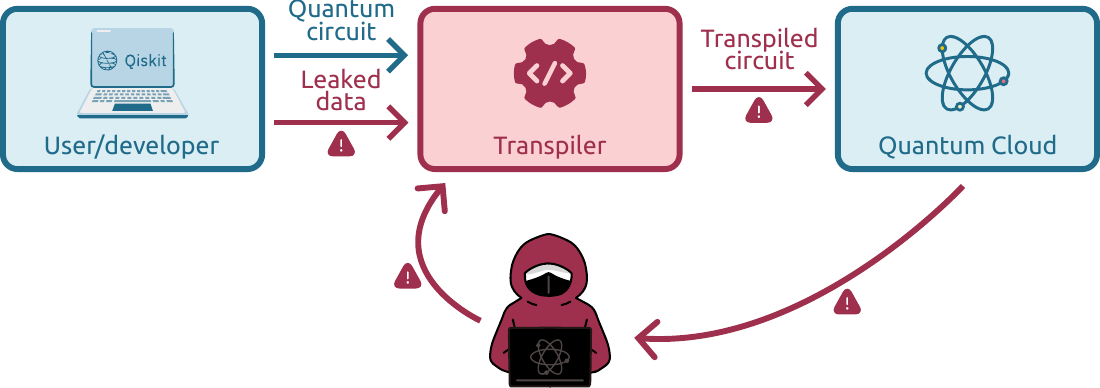}
\caption{Diagram of the threat model targeting confidentiality.}
\label{fig:diagram_confidentiality_threat}
\end{figure}

\paragraph{Scope} The objective is to protect the confidentiality of user data in quantum cloud workflows. Specifically, to prevent (classical) information leakage from the classical system used to prepare and submit quantum circuits to the cloud.

\paragraph{Attacker} The attacker can observe quantum jobs submitted to the cloud. This includes, in particular, the case of a malicious quantum cloud provider with full access to submitted circuits and execution metadata. Many quantum clouds provide a web interface to see submitted jobs and fetch results. However, it is a strong assumption that an attacker unrelated to the cloud provider has full access to submitted circuits, as these web interfaces require authentication.

\paragraph{Threats and threat vector} The attacker compromises the transpilation process. User's data is leaked and encoded into the final transpiled circuit sent to the quantum cloud. This exfiltration channel allows sensitive classical information to be encoded in gate choices, qubit layout, timing patterns of the final circuit, etc. After obtaining a copy of the transpiled circuit, the attacker can decode and extract the leaked data, violating user confidentiality.

\paragraph{Mitigation} The threat is detectable if the transpilation process is \emph{reproducible}. Full mitigation is achieved when users verify the determinism of the transpilation process before submission, ensuring that no information has been covertly injected. Techniques from reproducible builds in classical software can be adapted to quantum circuits to support this verification.

\subsection{Integrity}\label{sec:threat_model_integrity}

This threat model focuses on attacks that compromise the \emph{integrity} of quantum computations by subtly altering circuits during transpilation. Integrity in this context means that the transpiled circuit executed by the quantum backend corresponds to the original intended computation described by the source code, i.e., the initial quantum circuit. Preserving integrity is essential to ensure that the results produced by the quantum computer are trustworthy and scientifically meaningful. If the integrity is violated, the computation may yield degraded or incorrect results, undermining the usefulness of the quantum computer for its users. Maintaining integrity is therefore in the interest of both users and providers: users rely on correct execution for their applications and research, while providers depend on reliable results to build trust in their quantum platforms and services. In Section~\ref{sec:attacks-integrity}, it will be shown how an attacker can exploit public knowledge of the target backend to modify the transpilation output in a way that degrades the computation or produces entirely incorrect results, without direct access to the cloud infrastructure. Similarly to the confidentiality threat model, the scope of the attacks, assumptions, attacker capabilities, and threat vectors are specified precisely, so that it is clear how quantum reproducible builds can help mitigate these attacks.

\begin{figure}[ht]
\centering
\includegraphics[width=0.8\linewidth]{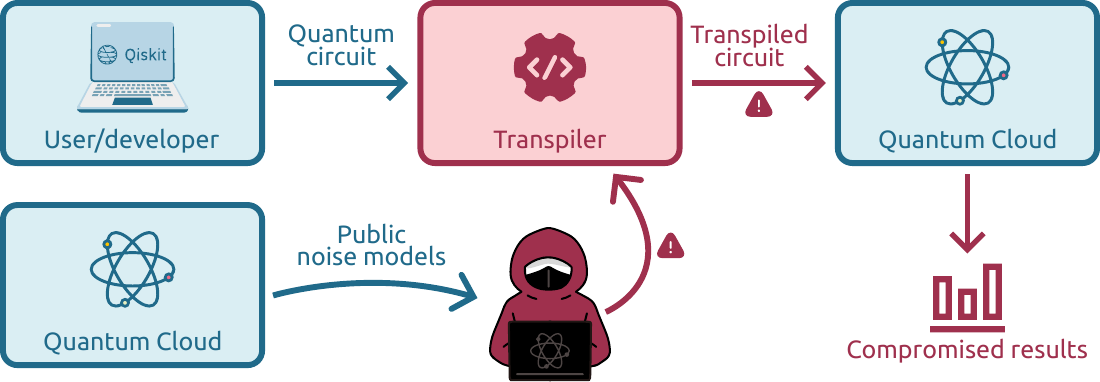}
\caption{Diagram of the threat model targeting integrity.}
\label{fig:diagram_integrity_threat}
\end{figure}

\paragraph{Scope} The goal is to preserve the integrity of quantum computation results. Specifically, to prevent malicious modifications to the transpiled circuits that could degrade accuracy or lead to entirely incorrect outcomes, relative to the original intended computation.

\paragraph{Attacker} The attacker does not have access to the user's submitted jobs or the quantum cloud infrastructure. However, they do have access to up-to-date noise models of the target backend used during transpilation, a realistic assumption since such data is typically public.

\paragraph{Threats and threat vector} The attacker tampers with the transpilation process by modifying gates, qubit assignments, parameter values, etc. These changes produce a quantum circuit that is no longer equivalent to the original, causing the computation to produce incorrect results, such as an altered probability distribution or an inaccurate expectation value of an observable. The attack may be subtle, targeting known hardware noise profiles to maximize its effect while remaining difficult to detect through output inspection alone.

\paragraph{Mitigation} As with the confidentiality threat, integrity violations can be detected, and ultimately prevented, if the transpilation process is reproducible. Users can compare the final circuit to a known-good version or rerun the process to check for determinism (e.g., generate the same transpiled circuit in two different computers and verify they are bit-by-bit identical). Ensuring reproducibility mitigates the risk of tampering and increases confidence in the correctness of submitted circuits.

\section{Data Leakage Using Non-Reproducible Quantum Circuits}\label{sec:attacks-confidentiality}

This section presents three examples that target different stages of the transpilation process, each compromising the confidentiality of end users' data as explained in Section~\ref{sec:threat_model_confidentiality}. The examples leverage the final transpiled circuits as covert exfiltration channels. While they only involve harmless data, such as text strings or small images, the same techniques could be adapted to extract sensitive information, including private cryptographic keys, authentication cookies, or other confidential documents. Although these examples cannot be directly exploited without additional weaknesses, such as inadequate access control to the quantum cloud, they underscore the importance of strong security and governance measures throughout the quantum software toolchain.

Importantly, small modifications to the transpiled circuit are harder to detect. An attacker's objective is therefore to minimize disturbances to the circuit while maximizing the amount of information encoded. These examples highlight a fundamental conceptual gap: currently, transpilation is not treated as a security-critical step, leaving it vulnerable to covert manipulation. Enforcing reproducible transpiled quantum circuits would make all of these attacks straightforward to detect and mitigate. Table~\ref{table:comparison_attacks_confidentiality} provides a brief comparison of the three examples.

\begin{table}[ht]
\centering
\begin{tabular}{lcl}
\toprule
Transpilation stage & Max leaked data size & Limitations \\
\midrule
Layout (Sec.~\ref{sec:layout-attack}) & $\lfloor\log_2 (\text{physical qubits})!\rfloor$  & Affects computation results \\
Init (Sec.~\ref{sec:init-attack}) & Unbounded & Requires, at least, one unused qubit \\
Scheduling (Sec.~\ref{sec:scheduling-attack}) & $6\times(\text{RZ gates})$ & Visual artifacts (see Fig.~\ref{fig:scheduling_attack_pi})\\
\bottomrule
\end{tabular}
\caption{Table comparing the three examples compromising confidentiality.}
\label{table:comparison_attacks_confidentiality}
\end{table}

\subsection{Modified layout stage}\label{sec:layout-attack}

The layout stage is the second step in the transpilation process (see Figure~\ref{fig:transpilation}). Only the unrolling of custom instructions and the translation of multi-qubit gates into single- and two-qubit gates occur earlier, during the init stage. The layout stage maps the virtual qubits of the quantum circuit onto the physical qubits available on the target backend. The layout stage example abuses this mapping as an exfiltration channel.

The trivial layout mapping is the following:
\begin{align}
\begin{split}
\mathcal{M}_\text{trivial}&: [n] \rightarrow [n]\\
\mathcal{M}_\text{trivial}(x) &\mapsto x
\end{split}
\end{align}

This mapping generally leads to bad quantum computation results because it does not take into account the connectivity of the real hardware, the error rates of quantum gates, etc. An optimal layout uses all available information from the quantum backend to obtain the best possible layout. For example, the optimal layout of a small 3-qubit circuit would map the virtual qubits to physical qubits that are all mutually connected and have the lowest noise levels at the time of execution, avoiding the need for additional SWAP gates, and ensuring the best possible results.

\begin{figure}[ht]
\centering
\includegraphics[width=0.7\linewidth]{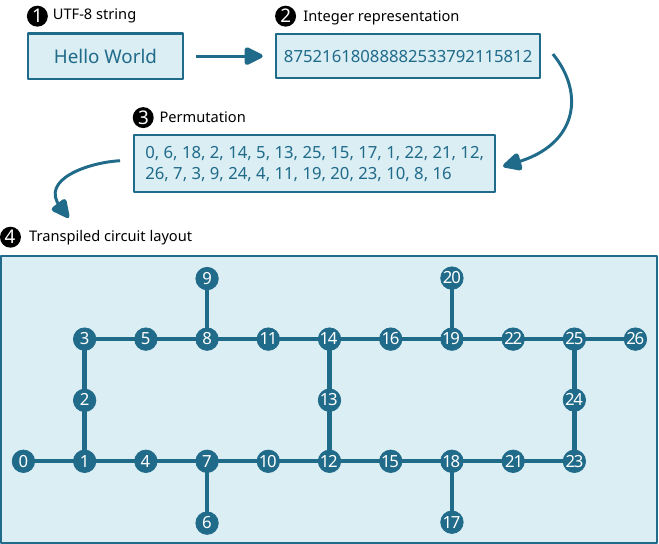}
\caption{\textbf{Example modifying the layout stage.}  First, the UTF-8 string \texttt{"Hello World"} is encoded as a large integer. Then, the integer is mapped to a permutation list using the Lehmer code~\cite{lehmer-code}. Finally, this permutation is used to define the mapping between virtual and physical qubits during the layout stage of the transpilation.}
\label{fig:layout_attack}
\end{figure}

A layout map can be seen as a permutation. For a quantum computer with $n$ qubits, there are $n!$ possible permutations or layout maps. This is because the first virtual qubit can be mapped to $n$ physical qubits, the second one to $n-1$, the third to $n-2$, and so on.

Information can be encoded using permutations. For a set of $n$ elements, it is possible to encode up to $\lfloor\log_2 n!\rfloor$ bits of information. Encoding and decoding can be implemented efficiently using e.g., the Lehmer code~\cite{lehmer-code}.

The modified layout stage proceeds as follows. First, the data to be leaked is converted into an integer using a bytes-to-integer mapping. Second, this integer is transformed into a permutation via a Lehmer code. Third, this permutation is applied during the layout stage of the transpilation. The resulting transpiled circuit, which is sent to the quantum cloud, encodes this mapping. This allows the leaked data to be recovered from the payload available to the attacker through the quantum cloud, as described in the threat model.

The amount of data that can be exfiltrated in the modified layout stage is bounded by the number of physical qubits. A quantum backend with 127 physical qubits\footnote{At the time of writing this paper, the smallest quantum backend available in IBM's Quantum Open Plan, \href{https://quantum.cloud.ibm.com/computers?system=ibm_brisbane}{\texttt{ibm\_brisbane}}, already has 127 qubits.} permits encoding up to 88 bytes of data. While 88 bytes may appear small, it is sufficient to exfiltrate secrets such as an Ed25519 private key, which can be used for authentication in the SSH protocol~\cite{openssh-ssh-authentication}, or a secp256k1 ECDSA private key, which can be used to authorize transactions from a Bitcoin wallet~\cite{bitcoin-secp256k1}. A small example using a backend with only 27 physical qubits is shown in Figure~\ref{fig:layout_attack}.

A limitation of the modified layout stage is that it leads to noticeably bad results in the quantum computations due to using a suboptimal layout that does not take into account the connectivity of the physical qubits.

\subsection{Modified init stage}\label{sec:init-attack}

The init stage is the first step in the transpilation process (see Figure~\ref{fig:transpilation}). In this stage, custom instructions and gates acting on three or more qubits are decomposed into single- and two-qubit gates, since subsequent stages of the transpilation assume circuits in this form.

The modified init stage avoids the limitation of the modified layout stage, namely that it can alter the quantum computation's results, and demonstrates that by manipulating only the init stage it is possible to leak arbitrarily large amounts of data.

\begin{figure}[ht]
\centering
\includegraphics[width=0.7\linewidth]{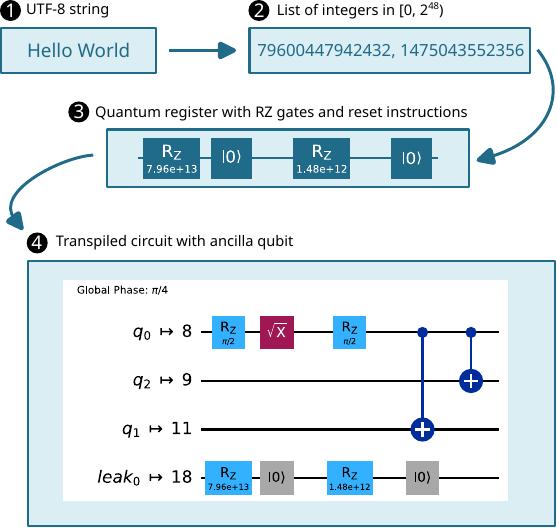}
\caption{\textbf{Example modifying the init stage.} First, the UTF-8 string \texttt{"Hello Word"} is encoded as a list of integers in the range $[0, 2^{48})$. Then, a new quantum register is created with RZ gates using those numbers as parameters. The rotation gates are surrounded by reset gates to protect them from modifications in later stages of the transpilation. Finally, the quantum register is appended to the quantum circuit using one additional qubit.}
\label{fig:init_attack}
\end{figure}

The modified init stage works in the following way. First, an auxiliary quantum register is added to the quantum circuit. Second, the data is encoded into integers in the range $[0, 2^{48})$. Third, rotation gates are inserted in the auxiliary qubits using the integers as rotation parameters. The rotation gates are surrounded by reset instructions to prevent later optimizations from combining them into a single gate. The choice of the integer range for the data encoding is due to implementation details of Qiskit (a detailed explanation can be read in the supplemental material; see \urlcolornameref{sec:code-data}).

A limitation of the modified init stage is that it requires that the input quantum circuit is, at least, one virtual qubit smaller that the total number of physical qubits available on the quantum backend. A small example of this modified transpilation stage is shown in Figure~\ref{fig:init_attack}.

\subsection{Modified scheduling stage}\label{sec:scheduling-attack}

The scheduling stage is the last stage in the transpilation process and it adds delay instructions to the quantum circuit to account for any timing constraints on the target backend. This is the perfect place to modify the final transpiled circuit and encode information into the already existing gates since no later stages will modify them.

\begin{figure}[H]
\centering
\includegraphics[width=0.7\linewidth]{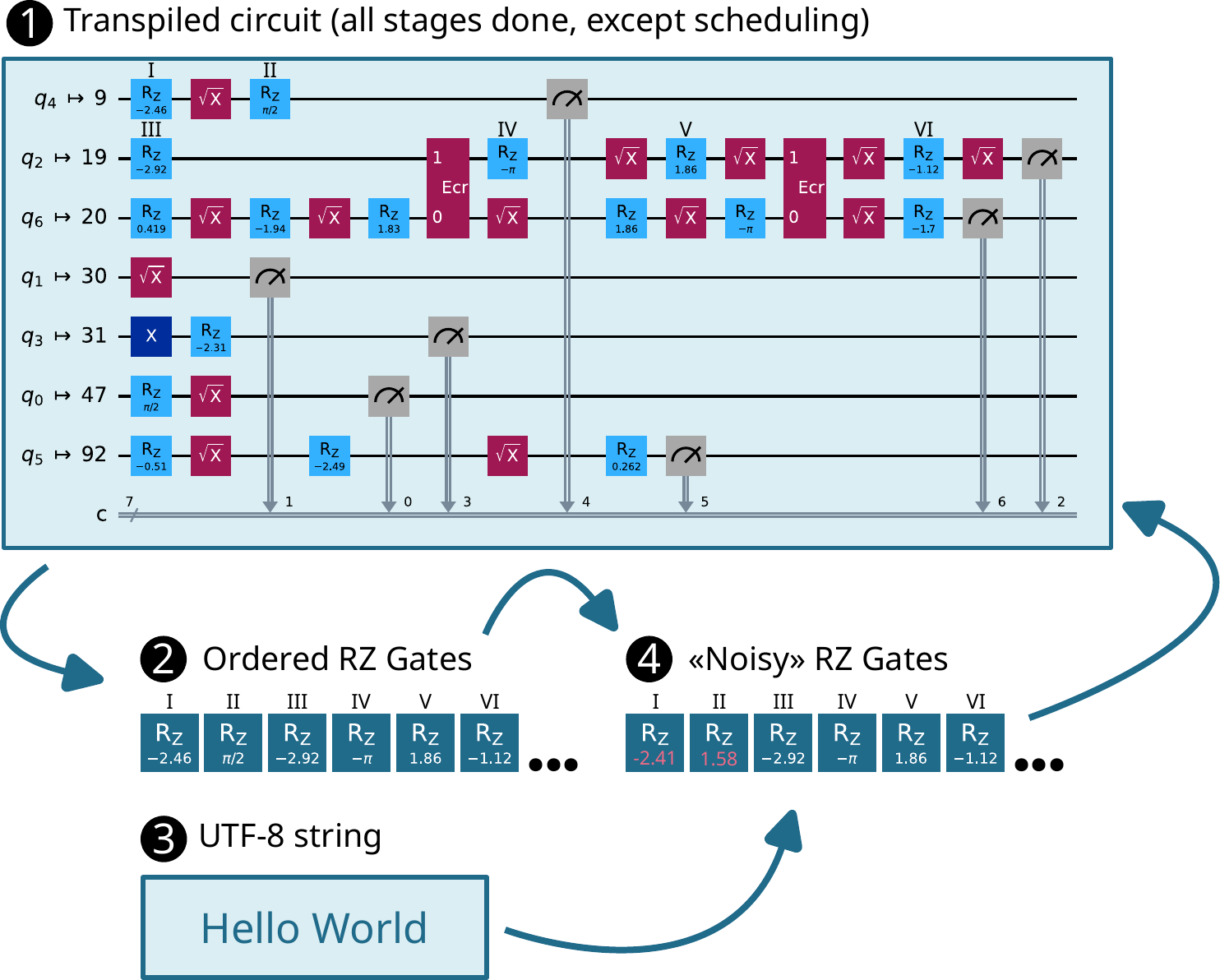}
\caption{\textbf{Example modifying the scheduling stage.} The RZ rotation gates (shown in light blue) are extracted from the transpiled quantum circuit and sorted in a predefined order: first, all gates on the physical qubit with the lowest index (from left to right), then those on the next qubit, and so on (as indicated by the Roman numerals in the figure). Each gate angle is then slightly modified to encode the data to be leaked. In this example, the 11-byte string \texttt{"Hello Word"} is encoded into the first two rotation gates, with the modified angles shown in red. These modified gates replace the original ones in the circuit. On real quantum backends, noisier RZ gates allow more information to be covertly embedded without altering the observable computation results. In current devices, up to 6 bytes can be encoded per gate; encoding up to 4 bytes per gate produces circuits visually indistinguishable from the originals, except for gates that contain multiples or fractions of $\pi$ (see Figure~\ref{fig:scheduling_attack_pi}).}
\label{fig:scheduling_attack}
\end{figure}

Both examples modifying the layout and the init stage introduce noticeable changes in the quantum artifacts: the modified layout changes the mapping between virtual and physical qubits, affecting the results of the quantum computations; the modified init introduces an ancilla qubit with useless rotations separated by reset instructions. The last of our examples targeting confidentiality demonstrates a stealthier modification that neither requires an additional ancilla qubit nor affects the computation results, at least on current noisy quantum computers. A diagram of the modified transpilation stage using a backend with 127 physical qubits is shown in Figure \ref{fig:scheduling_attack}.

\begin{figure}[H]
\centering
\begin{subfigure}{0.44\textwidth}
\includegraphics[width=\linewidth]{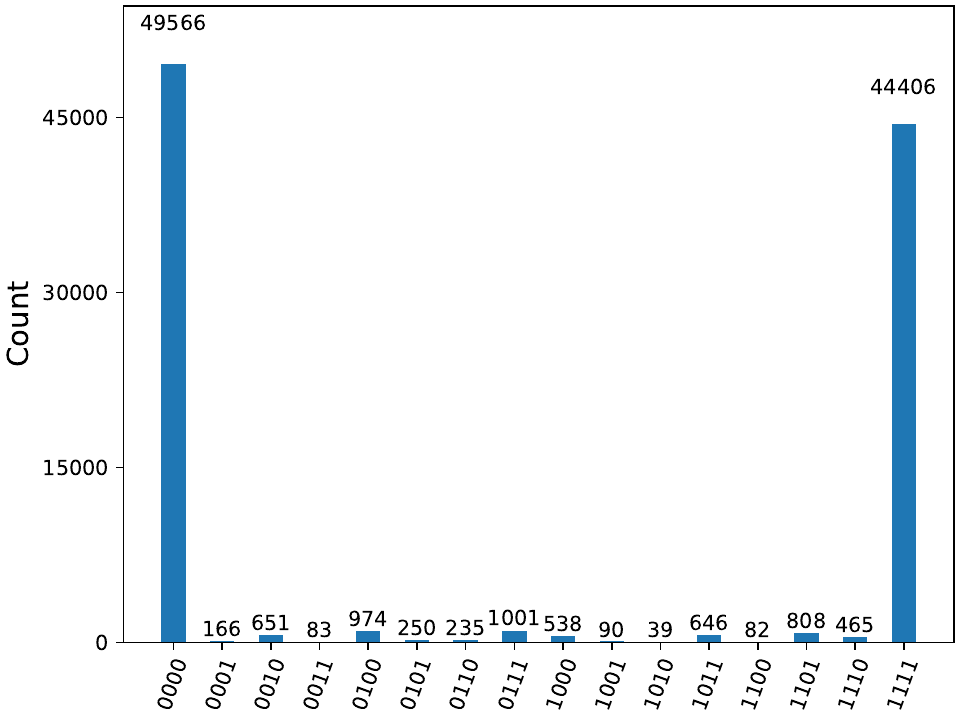}
\caption{Genuine transpiled circuit}
\label{fig:scheduling_attack_unmodified}
\end{subfigure}
\begin{subfigure}{0.44\textwidth}
\includegraphics[width=\linewidth]{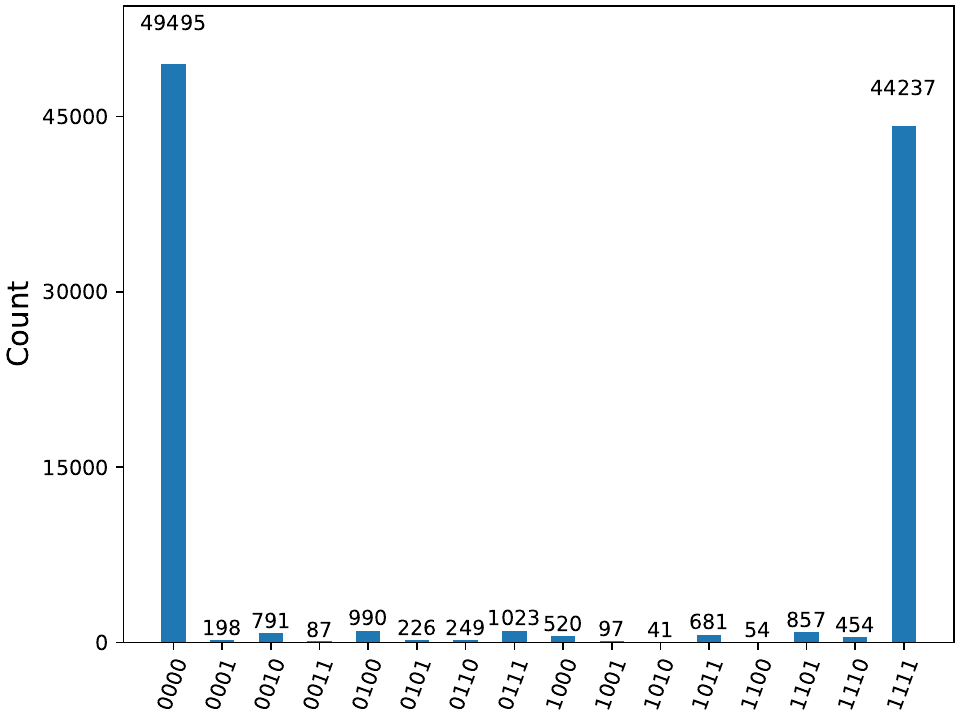}
\caption{Leaky transpiled circuit}
\label{fig:scheduling_attack_modified}
\end{subfigure}
\caption{Output distributions obtained after running on the real IBM quantum backend \href{https://quantum.ibm.com/services/resources?system=ibm_torino}{\texttt{ibm\_torino}} the 4-qubit GHZ circuit $10^5$ times. On the left, the genuine transpiled circuit. On the right, the leaky circuit with 66 random bytes encoded into its 11 rotation gates. Both circuits are indistinguishable with a Hellinger fidelity~\cite{Bengtsson_Zyczkowski_2006} of 0.9999. The results correspond to job ID \texttt{d33u85t0qhlc73cqu3dg} (15.09.2025).}
\label{fig:scheduling-attack-ibm-real-backend}
\end{figure}

The modified scheduling stage proceeds as follows. First, the circuit after the optimization stage (see Figure~\ref{fig:transpilation}) is analyzed and all the RZ rotation gates are extracted and sorted. Second, the data to be exfiltrated is split in blocks of 6 bytes (details about this particular choice can be read in the supplemental material; see \urlcolornameref{sec:code-data}). Third, these blocks are used to replace the 6 least significant bytes of each rotation angle. This example exploits current limitations of noisy quantum hardware. In particular, that small changes in the angles of the rotation gates are undetectable by observing the results of the quantum computation as shown in Figure~\ref{fig:scheduling-attack-ibm-real-backend}.

\begin{figure}[ht]
\centering
\includegraphics[width=0.5\linewidth]{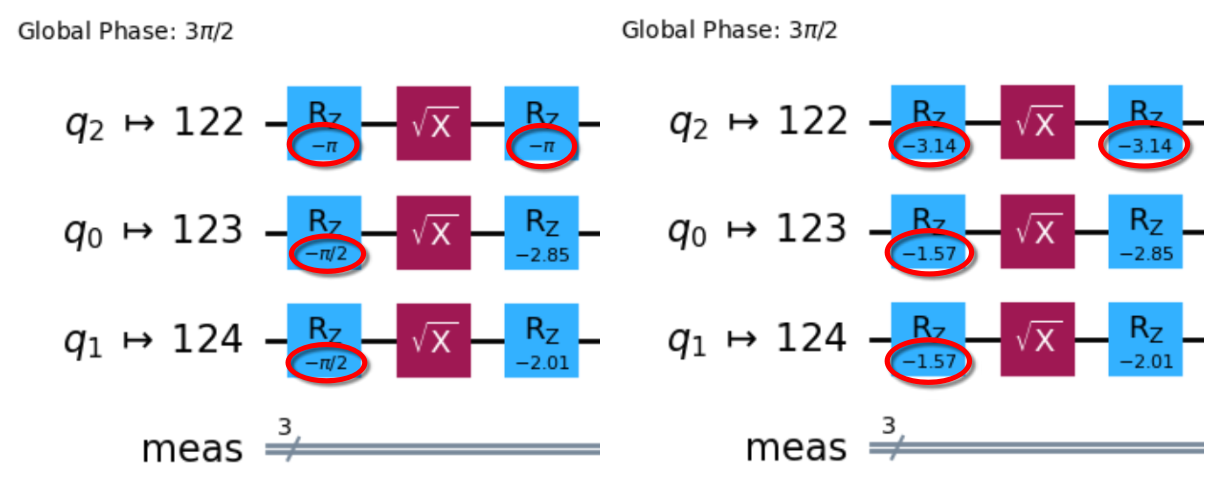}
\caption{Drawings of quantum circuits containing rotation gates. On the left the original transpiled circuit. On the right the modified transpiled circuit with the angles slightly altered by the modified scheduling stage changing the 4 least significant bytes. Note that the parameters no longer appear as multiples or fractions of $\pi$. This is caused by the function \href{https://github.com/Qiskit/qiskit/blob/9092ee782a413d22a1bc6cda248c18b2dc51ddba/qiskit/circuit/tools/pi_check.py}{\texttt{pi\_check()}}, which only draws $\pi$ for multiples or fractions within $10^{-9}$, but the modified transpilation stage introduces larger deviations. It is possible to avoid this visual artifact by encoding only 2 bytes per gate.}
\label{fig:scheduling_attack_pi}
\end{figure}

A limitation of the modified scheduling stage is that it produces small visual artifacts when the transpiled circuits are drawn as shown in Figure~\ref{fig:scheduling_attack_pi}. In addition, contrary to the modified init stage, the amount of information that can be exfiltrated is again bounded, in this case by the number of available RZ rotation gates in the transpiled circuit after the optimization stage.

\section{Compromised Integrity Using Non-Reproducible Quantum Circuits}\label{sec:attacks-integrity}

This section introduces two distinct examples that follow the threat model described in Section~\ref{sec:threat_model_integrity}, in which the attacker aims to compromise the integrity of quantum computations. These examples demonstrate that even minimal modifications to transpiled circuits can significantly degrade performance or lead to entirely incorrect outcomes. Integrity is essential for trusting the results from a quantum computer.

\subsection{GHZ circuit}
Benchmarking quantum computers is a non-trivial task, and arguably, it is not possible to design a single quantum circuit that captures all the features one might wish to evaluate. However, a very simple and effective circuit to test a quantum computer's ability to prepare a globally entangled state\footnote{An entangled state is a joint quantum state of multiple systems that cannot be written as a product of individual states, meaning the systems exhibit correlations that have no classical counterpart~\cite{NieChu00}.} is the GHZ circuit.

\begin{figure}[ht]
\centering
\begin{subfigure}{0.4\textwidth}
\includegraphics[width=\linewidth]{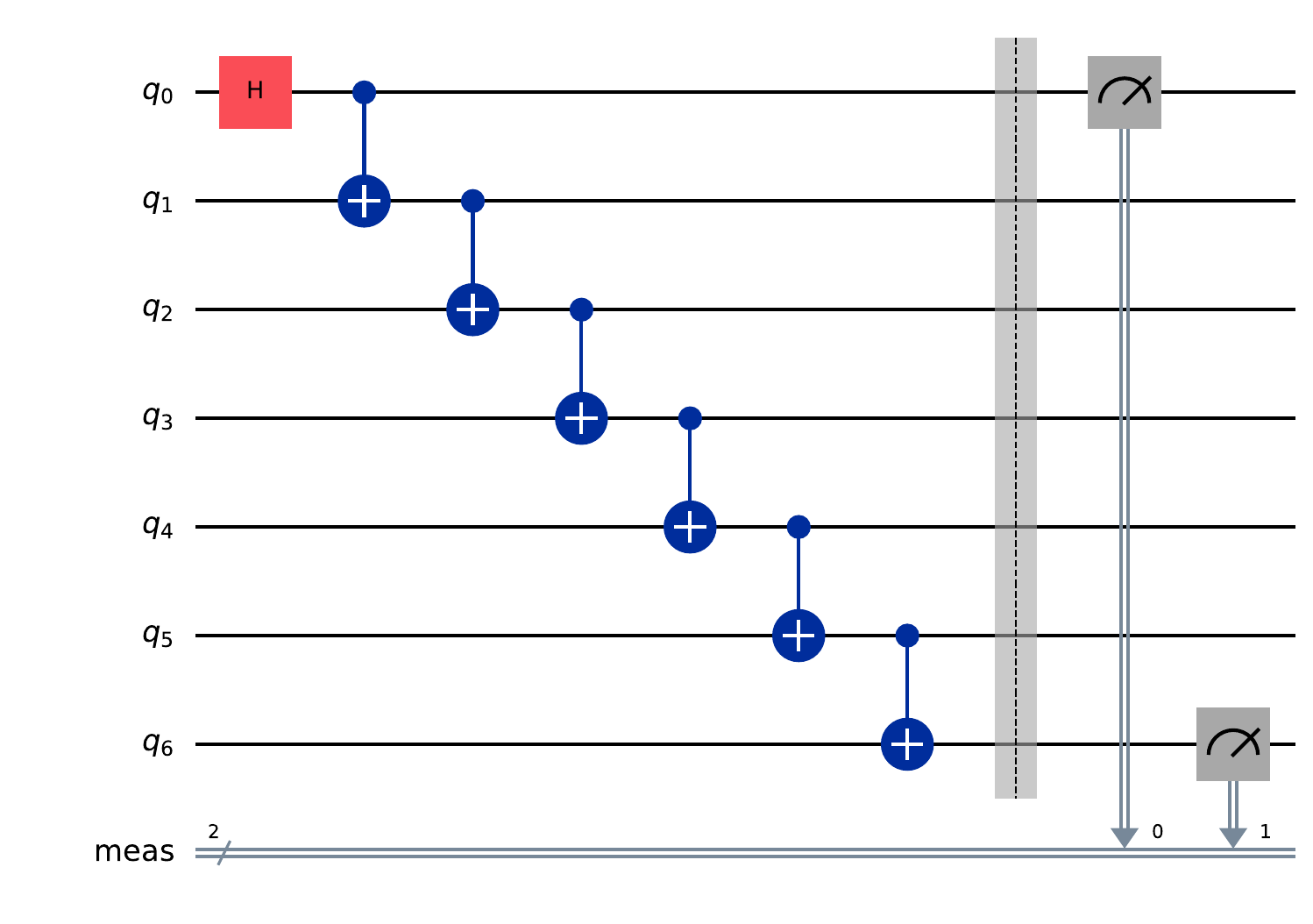}
\caption{7-qubit GHZ circuit}
\label{fig:7-qubit-ghz}
\end{subfigure}
\begin{subfigure}{0.4\textwidth}
\includegraphics[width=\linewidth]{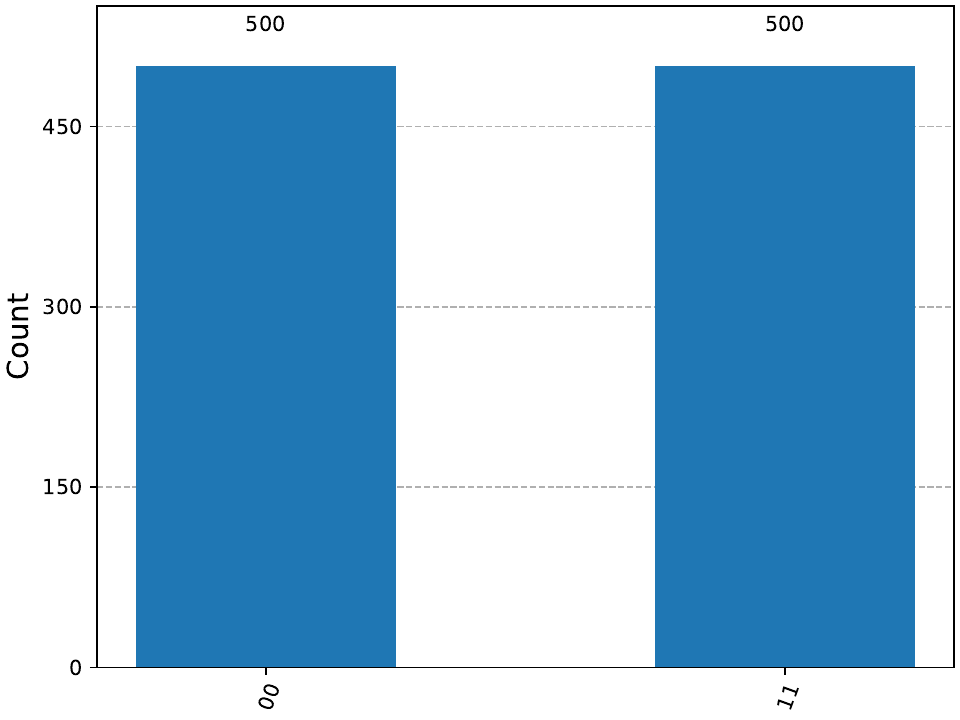}
\caption{Expected output distribution}
\label{fig:7-qubit-ghz-output}
\end{subfigure}
\caption{On the left, a 7-qubit GHZ circuit with two measurements on the first and last qubits. On the right, the expected output distribution from an ideal quantum computer.}
\label{fig:7-qubit-ghz-circuit}
\end{figure}

Ideally, all qubits should be measured at the end of the GHZ circuit. However, to simplify the resulting histograms, this example measures only the first and last qubits as shown in Figure~\ref{fig:7-qubit-ghz-circuit}. In the ideal case, the output should consist of 50\% \texttt{``00''} and 50\% \texttt{``11''}. Any other outcomes correspond to errors. While this simple setup does not allow to distinguish between different error sources, such as decoherence, faulty gates, or noisy measurements, it still provides a practical indicator of whether the intended entangled correlations are preserved.

\begin{figure}[H]
\centering
\begin{subfigure}{0.49\textwidth}
\includegraphics[width=\linewidth]{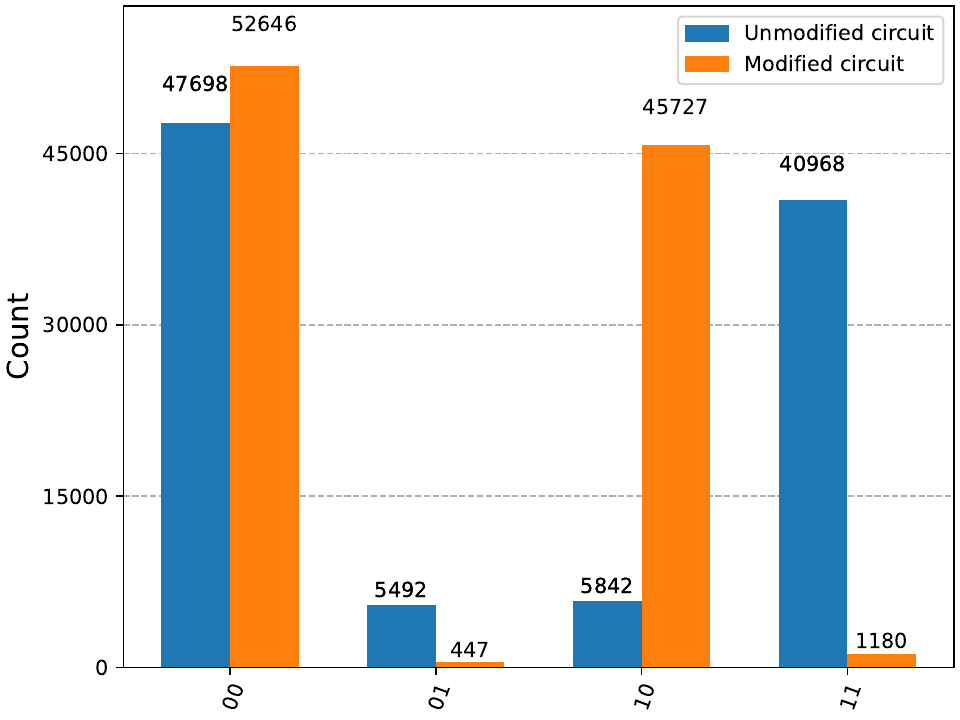}
\caption{Reset instruction in last qubit}
\label{fig:distribution_genuine_circuit}
\end{subfigure}
\begin{subfigure}{0.49\textwidth}
\includegraphics[width=\linewidth]{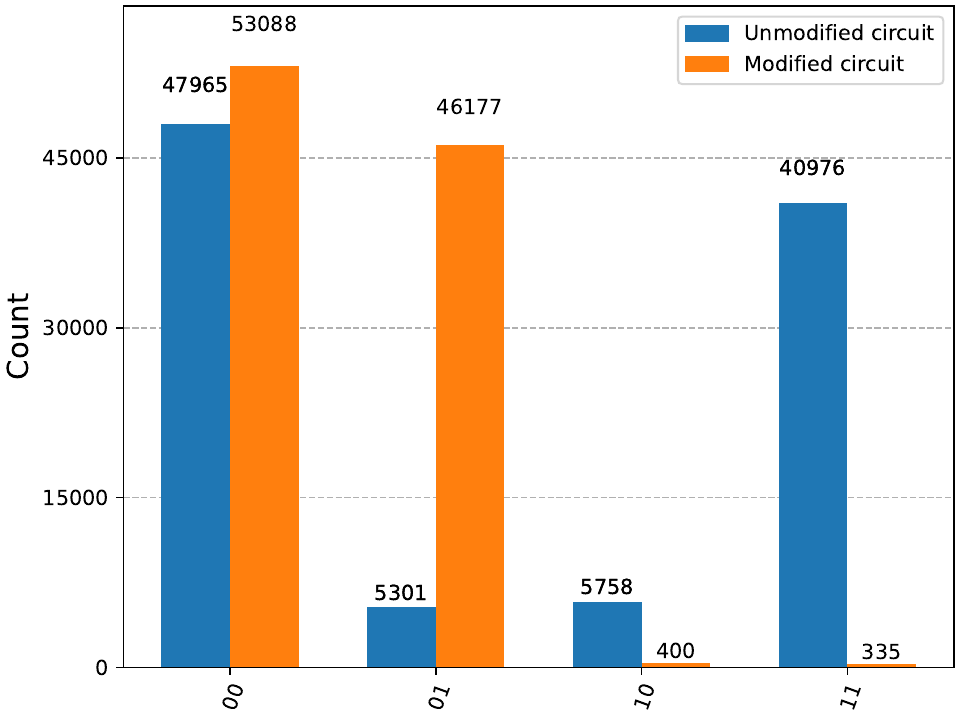}
\caption{Last CZ gate modified}
\label{fig:distribution_leaky_circuit}
\end{subfigure}
\caption{Output distributions of the unmodified and modified 7-qubit GHZ circuits obtained running on the real IBM quantum backend \href{https://quantum.ibm.com/services/resources?system=ibm_torino}{\texttt{ibm\_torino}} with $10^5$ shots. On the left, the modified circuit contains a reset instruction on the last (virtual) qubit before it is measured. On the right, the last CZ gate was modified to avoid entangling the last qubit. The results correspond to job IDs \texttt{d317rvhmc66s738e14c0} (11.09.2025) (left) and  \texttt{d30p6bbnfo5s73bhuq9g} (10.09.2025) (right).}
\label{fig:integrity_attack_ghz}
\end{figure}

It is possible to apply \emph{minimal} alterations to the transpiled circuit that drastically change the quantum computation results. Here, minimal refers to adding or removing a single instruction, or modifying an existing one. For example, inserting a single \texttt{reset} instruction before one of the measurements destroys the correlations and produces entirely different outputs. Similarly, changing the qubits on which a gate acts can significantly alter the computation results. Figure~\ref{fig:integrity_attack_ghz} illustrates two small modifications to the transpiled circuit that degrade the benchmarking results. A detailed version of this example is provided in the supplemental material (see \urlcolornameref{sec:code-data}).

\subsection{Grover's algorithm}

Grover's algorithm is a widely studied quantum algorithm that provides a quadratic speedup for solving unstructured search problems. Its output corresponds to a set of ``marked'' solutions, found with high probability by amplifying their amplitude through a sequence of Grover iterations. The full Grover circuit includes a quantum oracle and a diffusion operator, which is executed one or more times depending on the size of the search space and the number of marked solutions. Grover's algorithm is a useful test case for assessing the integrity of quantum computations since it is very sensible to small amounts of noise~\cite{Ijaz2023} (or malicious modifications).

\begin{figure}[H]
\centering
\begin{subfigure}{0.49\textwidth}
\includegraphics[width=\linewidth]{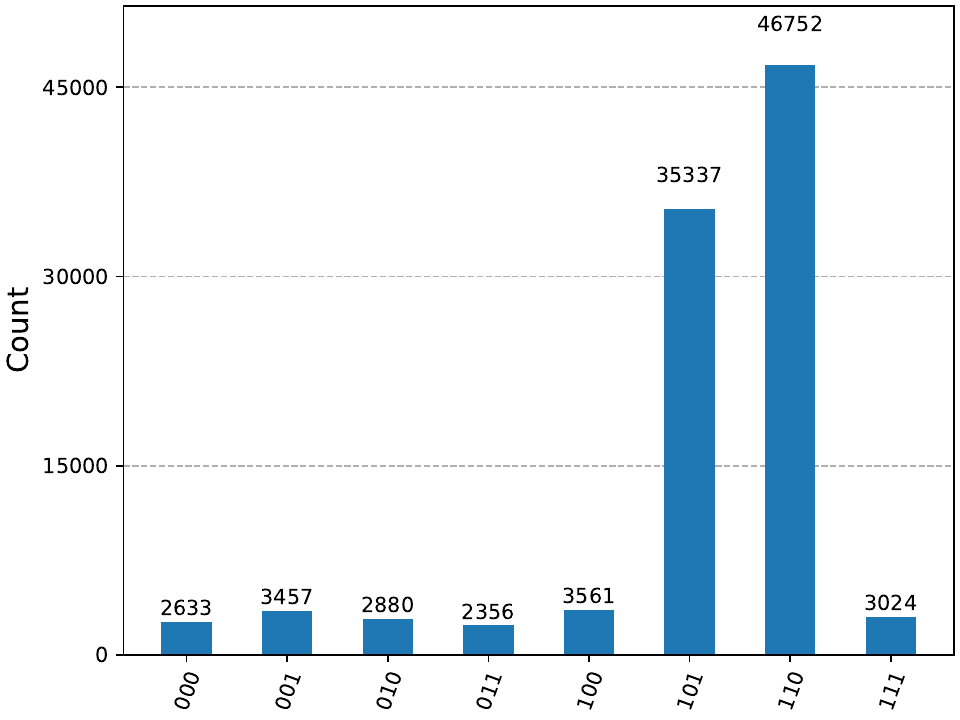}
\caption{Unmodified Grover transpiled circuit}
\label{fig:grover_unmodified}
\end{subfigure}
\begin{subfigure}{0.49\textwidth}
\includegraphics[width=\linewidth]{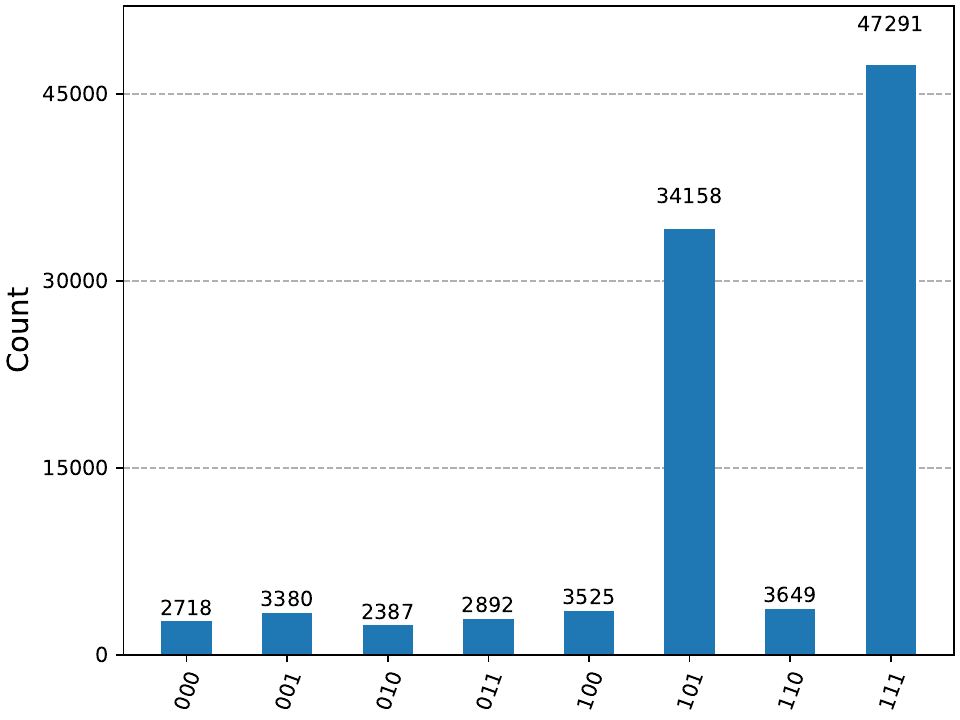}
\caption{Modified Grover transpiled circuit}
\label{fig:grover_modified}
\end{subfigure}
\caption{Output distributions of a full Grover circuit with an oracle marking states $\ket{101}$ and $\ket{110}$. The circuits were executed on IBM's 133-qubit backend \href{https://quantum.ibm.com/services/resources?system=ibm_torino}{\texttt{ibm\_torino}} $10^5$ times. The distribution on the left, with two clear peaks for ``101'' and ``110'', corresponds to the genuine transpiled circuit. The distribution on the right, on the other hand, shows incorrect results with a peak on ``111'' instead of ``110'' and it is the result of modifying the target qubit of a single CZ gate. The results correspond to job ID \texttt{d31g8nhmc66s738e961g} (11.09.2025).}
\label{fig:integrity_attacks_grover}
\end{figure}

Although it is not possible to run Grover's algorithm on current quantum computers to solve practical problems, it is possible to run the algorithm on small toy examples. In Figure~\ref{fig:integrity_attacks_grover}, a 3-qubit Grover's circuit is executed on a target quantum computer with 133 physical qubits to find the marked elements \texttt{``101''} and \texttt{``110''} from the small 3-bit search space. The transpiled version of this Grover's circuit, despite only having 3 qubits and requiring just a single usage of the diffusion operator, already has a depth and size (number of gates) larger than 100.

The genuine transpiled circuit was altered with a minimal modification, changing the target qubit of a single Controlled-NOT gate contained in the circuit, in order to recover the wrong marked states with high probability. This illustrates how quantum computations built on superposition and interference can be particularly vulnerable to integrity violations, as they are highly sensitive to small amounts of noise~\cite{Ijaz2023}. A compromised transpiler could introduce minimal perturbations that silently invalidate the result, without triggering any noticeable changes in the quantum artifacts.

\section{Making Quantum Artifacts Reproducible}\label{sec:ibm-r-b}

Making non-reproducible software reproducible is a non-trivial task, but it is achievable, even for large and complex projects, as demonstrated by successful efforts in projects like the Chromium web browser~\cite{chromium-deterministic-builds}, the official Bitcoin wallet~\cite{bitcoin-deterministic-builds}, and the Android-based operating system GrapheneOS~\cite{grapheneos-reproducible-builds}. Applying the same principles to quantum artifacts is easier now than it will be in the future, as the different available frameworks grow in complexity and potentially introduce new sources of non-determinism.

The remainder of this section focuses on the Qiskit SDK~\cite{qiskit2024} since it is arguably the most mature and used quantum computing framework available at the moment. First, the main sources of non-determinism in Qiskit's transpilation process are identified. Second, a series of changes to the Qiskit SDK are proposed to guide the framework towards producing reproducible transpiled quantum circuits.

\subsection{Sources of Non-Determinism}

The Qiskit SDK currently produces non-reproducible quantum artifacts primarily for two reasons: random seeds are employed during transpilation which are neither preserved nor propagated with the resulting artifacts; and the QPY format, not deterministic by design, is used to encode the final payload prior to submission to the quantum cloud. Unless reproducibility is established as an explicit design objective of the project and new tests are introduced to detect regressions in this regard, further sources of non-determinism may arise in future releases (see e.g.,~\cite{qiskit-bug-15035}).

\subsubsection{Random seeds}

Many of the problems that the transpiler has to solve are computationally expensive. For this reason, Qiskit uses randomized heuristic algorithms that require some initial randomness. By default, the pass manager obtains a fresh random seed every time it is initialized. In practice, this means that the exact same inputs, i.e., quantum circuit, level of optimization, passes and target backend, lead to different transpiled circuits.

\subsubsection{QPY encoding}

Before a transpiled quantum circuit is sent to a quantum cloud, it is encoded using the QPY format~\cite{QPY-format}. By design, QPY is a binary serialization format that aims to be cross-platform, Python version agnostic, and backwards compatible. It is meant to be used as a safe way to copy or save quantum circuits between different systems without loss of information and preserving the full Qiskit object structure. However, these full Python objects contain random strings (e.g., each circuit parameter contains a unique random UUID), counters (e.g., unnamed quantum circuits contain a counter index), memory addresses, etc. and are, therefore, not deterministic. This particular choice of encoding leads to non-reproducible quantum artifacts, i.e., an exact same transpiled quantum circuit can produce different final binary payloads.

\subsection{Proposed changes for Qiskit}

The following outlines the minimal set of changes required for the latest versions of the Qiskit SDK and Qiskit IBM Runtime\footnote{At the time of writing, the latest versions of Qiskit and Qiskit IBM Runtime are \href{https://github.com/Qiskit/qiskit/releases/tag/2.2.1}{v2.2.1} and \href{https://github.com/Qiskit/qiskit-ibm-runtime/releases/tag/0.42.0}{v0.42.0}, respectively.} to produce reproducible quantum artifacts.

\begin{enumerate}
    \item Implement a function that generates a \texttt{buildinfo} file containing all the necessary information for reproducibility. This file should include at least: the versions of packages and plugins used, random seeds set during transpilation, and relevant configuration details. With all these additional information, and the source code of the quantum circuit, anyone should be able to obtain identical final transpiled circuits.

    \item Replace the use of QPY to serialize the transpiled quantum circuits with a text-based OpenQASM~\cite{OpenQASM3} encoding. The QPY format, by design, has some goals that are not aligned with reproducible builds. In particular, the goal of preserving the full Python object might be interesting to make backups or share circuits between different systems, but it does not allow for reproducible payloads. Furthermore, many of the data structures included in the QPY payload are not relevant to the execution of the circuit on the quantum backend.
\end{enumerate}

The first change is relatively straightforward to implement and would have minimal impact on the rest of the Qiskit framework. The second change, however, represents a more fundamental shift in how the Qiskit Runtime IBM Client communicates with the backend, and would likely require corresponding updates on the server and quantum backend side. Beyond eliminating current and potential future sources of non-determinism, this change also reduces the amount of data transmitted from client to backend. While this may not be very relevant today, it could become increasingly important as the adoption of quantum computing continues to grow.

\section{Conclusion and Future Work}\label{sec:conclusion}

The well-established concept of reproducible builds has been extended to current quantum computing workflows. This new definition is motivated by examples that threaten both the confidentiality of quantum computer users and the integrity of quantum computation results. While these examples are not directly exploitable in practice without additional weaknesses, such as insufficient access control or insecure build environments, they highlight structural vulnerabilities that reproducible builds can help address.

The main sources of non-determinism in Qiskit were identified, and a minimal set of modifications was proposed to ensure that both Qiskit and Qiskit IBM Runtime can produce reproducible transpiled quantum circuits. These changes enable the Qiskit SDK to support reproducible quantum builds, thereby helping detect and mitigate the threats demonstrated in this paper.

While this work focuses on Qiskit, similar threat vectors are expected to exist in other quantum programming frameworks if their final artifacts are not reproducible. Future research could also explore how independent verifiers can be used to check the reproducibility of transpiled circuits, and how their results can enhance transparency and strengthen trust in emerging quantum cloud platforms.

\newpage

\section*{Acknowledgements}

This work was supported by the Swiss National Science Foundation Practice-to-Science Grant No 199084.

\section*{Code and Data Availability}\phantomsection\label{sec:code-data}

Implementations and more technical details of all the examples described in this paper are available as Jupyter notebooks in the GitHub repository \simpleicon{github}~\href{https://github.com/cryptohslu/reproducible-builds-quantum-computing}{cryptohslu/reproducible-builds-quantum-computing}. The three modified transpilation stages targeting confidentiality are also provided as Qiskit plugins on the Python Package Index (PyPI) \simpleicon{pypi}: \href{https://pypi.org/project/qiskit-leaky-layout}{qiskit-leaky-layout}, \href{https://pypi.org/project/qiskit-leaky-init}{qiskit-leaky-init}, and \href{https://pypi.org/project/qiskit-leaky-scheduling}{qiskit-leaky-scheduling}.

\bibliographystyle{quantum}
\bibliography{references}
\end{document}